\documentclass[fleqn,usenatbib]{mnras}

\usepackage{newtxtext,newtxmath}

\usepackage[T1]{fontenc}
\DeclareRobustCommand{\VAN}[3]{#2}
\let\VANthebibliography\thebibliography
\def\thebibliography{\DeclareRobustCommand{\VAN}[3]{##3}\VANthebibliography}

\usepackage{graphicx}	
\usepackage{amsmath}	
\usepackage{threeparttable}

\title[Kinematics and Stellar Population of R21]{A Comprehensive Look at PUDG-R21: Stellar Population and Kinematics of a Globular Cluster-Rich Ultra-Diffuse Galaxy in the Perseus Cluster}

\author[Levitskiy et al.]{
Arsen Levitskiy,$^{1}$\thanks{E-mail: arsenlv.1115@gmail.com} 
Duncan A. Forbes,$^{1}$
Jonah S. Gannon,$^{1}$
Anna Ferr\'e-Mateu,$^{2,3,1}$
Aaron J. Romanowsky,$^{4,5}$
\newauthor{\,\,Jean P. Brodie,$^{1,6}$
Warrick J. Couch,$^{1}$
Lydia Haacke$^{1}$}
\\
\\
$^{1}$\,Center for Astrophysics and Supercomputing, Swinburne University, John Street, Hawthorn VIC 3122, Australia\\
$^{2}$\,Instituto de Astrof\'isica de Canarias, Av. Via Lactea s/n, E38205 La Laguna, Spain\\
$^{3}$\,Departamento de Astrof\'isica, Universidad de La Laguna, E-38200, La Laguna, Tenerife, Spain\\
$^{4}$\,Department of Physics and Astronomy, San Jos\'e State University, One Washington Square, San Jose, CA 95192, USA\\
$^{5}$\,Department of Astronomy and Astrophysics, University of California Santa Cruz, 1156 High Street, Santa Cruz, CA 95064, USA \\
$^{6}$\,University of California Observatories, 1156 High Street, Santa Cruz, CA 95064, USA
}

\date{Accepted XXX. Received YYY; in original form ZZZ}

\pubyear{\the\year{}}

\begin{document}
\label{firstpage}
\pagerange{\pageref{firstpage}--\pageref{lastpage}}
\maketitle

\begin{abstract}
We present the analysis of the stellar populations and kinematics of the globular cluster (GC) rich ultra-diffuse galaxy, PUDG-R21, using spectroscopic observations obtained with the Keck Cosmic Web Imager (KCWI). The recessional velocity is measured to be 5536\,$\pm$\,10 km\,s$^{\mathrm{-1}}$, confirming its association with the Perseus cluster. The galaxy exhibits mild rotation of 15.6\,$\pm$\,10 km\,s$^{\mathrm{-1}}$ and a stellar velocity dispersion of 19.4\,$\pm$\,3.5 km\,s$^{\mathrm{-1}}$ within the galaxy effective radius. From this, we infer a dynamical mass of M$_{\mathrm{dyn}}=9.3\pm3.3\times10^{8}$\,M$_{\odot}$. Based on a halo mass derived from PUDG-R21 GC counts, we find our dynamical mass is consistent with a cored dark matter profile. The integrated stellar population analysis reveals a predominantly old stellar population of $10.4\pm1.2$\,Gyr, with intermediate-low metallicity ([M/H]$=-0.64\pm0.12$\,dex) and elevated alpha abundances ([Mg/Fe]$=0.38\pm0.25$\,dex). The inferred star formation history suggests rapid stellar assembly, likely truncating prior to or during the galaxy's infall into the cluster at an early epoch ($\sim$10\,Gyr ago). The analysis of stellar population gradients (age and metallicity) indicates a flat profile out to one effective radius. 
Here, we consider the involvement of two star formation events, initially forming a large population of metal-poor globular clusters, and then the latter contributing to the more metal-enriched diffuse stellar body. The evidence of subsequent star formation suggests this galaxy is more like an extension of the classical dwarf population than the much discussed failed galaxy UDGs. 

\end{abstract}

\begin{keywords}
    {galaxies: evolution, galaxies: dwarfs, galaxies: stellar content, galaxies: kinematics and dynamics, techniques: spectroscopic}
\end{keywords}


\section{Introduction}\label{sec:intro}

The recognition of a class of galaxy, first formally defined as ultra diffuse galaxies (UDGs) by \citet{VanDokkum2015}, has lead to a resurgence of interest in diffuse, low surface brightness galaxies (LSBs). UDGs are characterized by their large effective radii ($R_{\mathrm{e}}\ge1.5$\,kpc) and low surface brightness ($\mu_{0,g}\ge24$\,mag\,arcsec$^{-2}$). 
Since 2015, substantial efforts have been made to study UDGs across different environments, revealing large populations in clusters \citep[e.g.][]{VanDokkum2015, Yagi2016, Janssens2017, Wittmann2017, Janssens2024}, groups \citep[e.g.][]{Forbes2019, Prole2019, Iodice2020, Lim2020, LaMarca2022}, and in the field \citep[e.g.][]{Leisman2017, Janowiecki2019, Zaritsky2021, Jones2023}. Studies of UDG stellar populations have found a subpopulation that exhibits intermediate-to-old ages, often with more extended star formation histories, and metallicities that follow the present mass-metallicity relation (MZR) \citep[e.g.][]{Anna2018, Anna2023, Buzzo2022, Buzzo2024}. The resemblance of these properties to those of classical dwarf galaxies sparked the idea of ``puffing up" such galaxies to explain the formation of these UDGs. This explanation involves either internal processes, such as strong supernova feedback \citep[e.g.][]{Dicintio2017} or high halo spins \citep[e.g.][]{Amorisco2016, Rong2017}, or external processes, such as galaxy mergers \citep[e.g.][]{Wright2021}, tidal heating, and ram pressure stripping \citep[e.g.][]{Yozin2015, Mistani2016, Carleton2019, Sales2020}, with some suggesting a combination of both processes \citep[e.g.][]{Jiang2019}. 

Nonetheless, a second subpopulation of UDGs has been found to exhibit a strikingly different set of properties, including early-quenched, very old stellar populations, high alpha abundances, and with extremely low metallicities that offset them from the present MZR of local dwarfs \citep[e.g.][]{Anna2018, Anna2023, Buzzo2022, Buzzo2024}. These galaxies are also found to have extraordinarily high globular cluster (GC) richness for their stellar masses, implying an unusually massive dark matter halo, as inferred from the GC-halo mass relation by \citet{Spitler2009, Burkert2020}. These properties earned them the name ``failed galaxies" \citep{VanDokkum2015, Peng2016, Danieli2022, Forbes2024}. The cause of their early quenching of these galaxies still remains a conundrum, with various explanations presented, including environmental quenching \citep{Yozin2015, Mistani2016, Carleton2019}, a cosmic sheet/filament interaction \citep{Pasha2023}, and feedback-based quenching \citep{Danieli2022}.

The Perseus Cluster has been the subject of several large photometric studies of LSBs/UDGs, incorporating imaging from HST \citep{Harris2020, Janssens2024}, WHT \citep{Wittmann2017}, Subaru \citep{Gannon2022}, and more recently, Euclid \citep{Marleau2025}. These studies have focused on the structural properties and colours of the dwarf galaxies, as well as their GC systems. Up to ten UDGs/LSBs in the Perseus cluster have been found to host large populations of GCs ($>$20), accounting for up to a few percent of the total stellar mass of the host galaxy \citep{Janssens2024}. In the latter, it was also found that the mean colours of these GCs closely resemble the host galaxy’s colours, suggesting a similar stellar population age and metallicity. 
These GC-rich UDGs therefore seem to favour the failed galaxy formation scenario, implying that they reside in overly massive dark matter halos with fast-and-early quenched stellar populations \citep[e.g.][]{Forbes2025}. Nonetheless, information regarding the galaxy kinematics and stellar populations necessitates deep spectroscopic follow-up observations. 

Given the faint nature of UDGs, spectroscopic studies remain costly due to the long observational integration times required. To date, a few dozen of targets have been spectroscopically examined in detail, generally finding them to have intermediate-to-old mass-weighted ages, $\langle t_{M} \rangle=8.3\pm3.3$\,Gyr, enhanced alpha abundances, $\langle$[Mg/Fe]$\rangle=+0.51\pm0.33$\,dex, and low  metallicities, $\langle$[M/H]$\rangle=-1.03 \pm0.37$\,dex \citep{Anna2018, Anna2023}. 
Despite this, two of the three spectroscopically investigated GC-rich UDGs in Perseus cluster show slightly more enhanced metallicities ([M/H] $=-0.4$\,dex and [M/H]$=-0.61$\,dex;  \citet{Anna2023}), differing from the metal-poor failed galaxy candidates \citep{Buzzo2022, Anna2023, Buzzo2024, Forbes2025}. While this could be a sign for a different formation pathway of these UDGs, a larger sample of GC-rich UDGs is needed to investigate this matter. 

In this work, we thus targeted one of the most GC-rich UDG observed in the Perseus cluster, PUDG-R21 (hereafter referred to as R21), which hosts 36$_{-8}^{+8}$ GC candidates \citep{Janssens2024}. Similar GC numbers, $\rm N_{GC}=37_{-15}^{+17}$, for R21 were derived in \citet{Li2024} using a more novel approach incorporating a hierarchical Bayesian point process model. For this study, we adopted the GC number with the smaller uncertainty. The galaxy has an effective radius $R_{\mathrm{e}}=2.16$\,kpc, with a $\mu _{0,\mathrm{F475W}}=24.17\pm0.06$\,mag arcsec$^{-2}$, classifying it as a UDG in \citet{Janssens2024}. 
Adding to the sample of spectroscopically studied UDGs in the Perseus cluster, we delve into the global kinematics properties in Section\,\ref{sec:kinematics} and stellar population properties in Section\,\ref{sec:stellarpop} of R21,  

In addition to the above, a promising direction lies in the radial properties of these galaxies. Stellar population gradients within galaxies provide crucial insights into their feedback mechanisms and dynamical histories. A handful of UDGs have been investigated for the presence of stellar population age and metallicity gradients. These include DF44 \citep{Villaume2022}, four UDGs/Nearly UDGs (NUDGes) from \citet{Anna2024}, and FCC\,224 in the Fornax cluster \citep{Buzzo2025a}, all of which exhibit flat age gradients and, more surprisingly, flat-to-positive metallicity gradients. The latter is contrary to the predictions of simulations \citep[e.g.][]{Cardona2023, Benavides2024}. Flat age gradients are expected given the relatively short star formation timescales \citep[e.g.][]{Riggs2024}, but the flat-to-rising metallicity gradients are puzzling.  
In Section\,\ref{sec:stellargrad}, we utilize both spectroscopic and broadband data to recover the colour and stellar population gradients of R21. 
Throughout this study, we adopt a $\Lambda$CDM cosmology with $H_0 = 70 \; \rm km  \; s^{\mathrm{-1}} \; Mpc^{\mathrm{-1}}$ and a distance $d=75$\,Mpc to the Perseus cluster. For R21, we adopt an $R_{\mathrm{e}}$ of 2.16\,kpc, with galaxy axial ratio (b/a)$=0.85$ and Position Angle (PA) $=90^{\circ}$. 

\section{Observations}\label{sec:obs}

\subsection{Data Acquisition and Reduction}\label{subsec:data}

This work utilizes spectroscopic data obtained from the Keck Cosmic Web Imager (KCWI) \citep{Morrissey2018} as part of observations conducted on the nights of 2023 October 23 (program: W277, PI: Gannon) and 2024 November 6 (program: W283, PI: Gannon). The KCWI configuration for the first night used the medium slicer with the BL grating set with a central wavelength of 4550\,\AA\, on the blue arm and the RH3 grating with a central wavelength of 8600\,\AA\, on the red arm. The second night used the medium slicer with the BH3 grating set with a central wavelength of 5110\,\AA\, for the blue arm and the RH2 grating with the central wavelength of 6750\,\AA\, for the red arm. Data from the red arm with the RH2 and RH3 gratings were not used in this current work due to low signal-to-noise ratios (S/N). A summary of the number of exposures and total integration time for each night is provided in Table\,\ref{tbl:kcwi_obs}. The spectral coverage of the BL grating spectra ranged from 3630 to 5650\,\AA, and an instrumental resolution of $\sigma_{\rm inst}=67$\,km\,s$^{\mathrm{-1}}$. For spectra obtained with the BH3 grating, the spectral range was 4868 to 5347\,\AA, and a spectral resolution of $\sigma_{\rm inst}=13$ km\,s$^{\mathrm{-1}}$.  Throughout the two nights of observation, the skies were dark with seeing conditions ranging from 1.25$\arcsec$ to 1.5$\arcsec$.

The data reduction was performed using the standard KCWI pipeline, with the exception of the pipeline’s sky subtraction procedure. Additional trimming and corrections for flat-fielding were applied following the procedures outlined in \citet{Gannon2020}, before the data were mosaic stacked using the open source code $\rm Montage$ \citep{Jacob2010}. As the target covers an approximate area of 16\arcsec$\,\times$\,20\arcsec (30\,$\times$\,44 spaxels), an on-chip global sky spectrum was extracted from the empty sky regions within the slicer for custom background subtraction. 

\renewcommand{\arraystretch}{1.1}%
\begin{table}
    \centering
	\caption{A summary of the KCWI gratings, central wavelength, program ID, date and exposure times for observations of R21.}
	\label{tbl:kcwi_obs}
	\begin{tabular}{ccc} 
		\hline
		Central $\lambda$\, and Grating & Program ID and Date & Exposures\\
		\hline
        4550 / BL & W277 (2023 Oct 23) &  8 $\times$ 1320s \\
        8600 / RH3 & W277 (2023 Oct 23) &  32 $\times$ 300s \\
		\hline
        5110 / BH3 & W285 (2024 Nov 6) &  5 $\times$ 1620s\\
        6750 / RH2 & W285 (2024 Nov 6) & 13 $\times$ 600s\\
        \hline
	\end{tabular}
\end{table}

\begin{figure}
\centering
\includegraphics[width=\columnwidth]{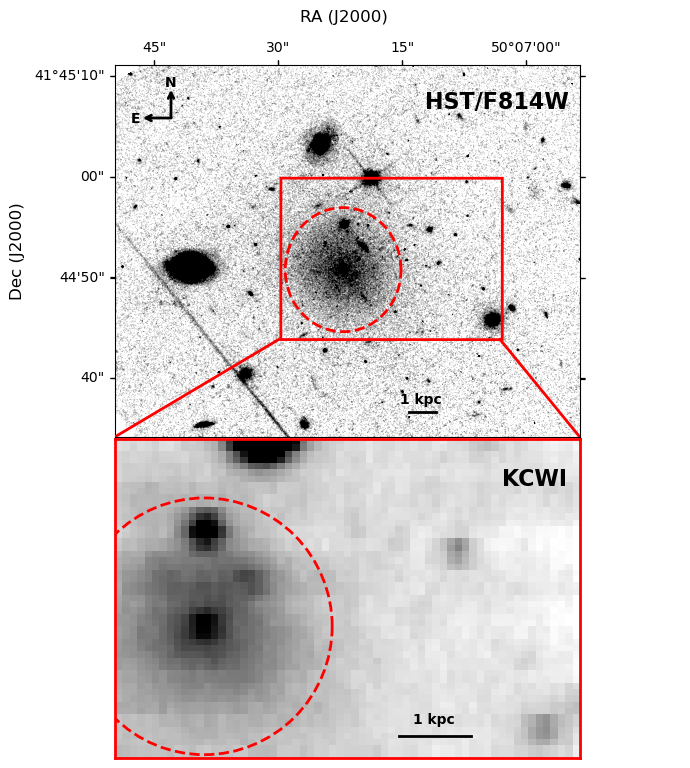}
\caption{HST and KCWI observations of R21. Top panel: HST image of R21 captured using the F814W filter on the ACS. Red rectangle shows the field of view (FoV) of R21 as observed by KCWI. 
The red dashed circular region marks the effective radius at 2.16\,kpc. Both images have North up and East left and show a 1\,kpc scale bar at a distance of 75\,Mpc.}
\label{fig:FoV}
\end{figure}

               
We additionally utilized HST images from the Program for Imaging of the Perseus Cluster of Galaxies (PIPER) \citep{Harris2020}. This dataset consists of two images taken with the Advanced Camera for Surveys (ACS/WFC) using the F475W and F814W filters. All the images have a pixel size of 50\,mas with all the magnitudes calibrated to the VEGA magnitude system. We present the reduced KCWI white light image in Figure\,\ref{fig:FoV}, along with its full field of view in comparison to the HST/ACS F814W image. All spectra and photometry were corrected for foreground reddening, assuming the total extinction of $A_{\mathrm{v}}=0.46$ based on the updated dust map of the Milky Way by \citet{Schlafly2011}.

\vspace{4mm}

\section{Stellar Kinematics}\label{sec:kinematics}

One of the benefits of having two sets of observations with different blue gratings and thus different spectral resolutions is their complementarity. The high signal-to-noise ratio (S/N) of the low-resolution BL grating spectra provides a good measurement for the recessional velocity (RV) of R21 as well as a more extended measurement of the stellar bulk rotation (Section\,\ref{subsec:rv+rotation}). Examination of the physical properties of its stellar population (Section\,\ref{sec:stellarpop}) also utilized the BL spectrum. Despite its lower S/N, the finer resolution of the BH3 grating is sufficient to accurately measure the global velocity dispersion of R21 (Section\,\ref{subsec:vdisp}).


        

\subsection{Recessional Velocities and Stellar Rotation}\label{subsec:rv+rotation}

We first identified all the sources' spaxels using Photutils\footnote{https://photutils.readthedocs.io} source detection and segmentation functions on the white light collapsed KCWI BL image, adopting a 2$\sigma$ threshold above the local background. 
We extracted the integrated spectrum of R21 using a global aperture of 1\,R$\rm _{e}$ (red dashed circle in Figure\,\ref{fig:FoV}) around the galaxy (see later Section\,\ref{sec:stellarpop} for the global BL spectrum). For measuring the systemic velocity, we utilized the high resolution stellar template library from \citet{Coelho2014}, and conducted the fits with the full-spectral fitting code {\sc pPXF} \citep{Capperllari2017}. To ensure the consistency of our outputs, we tested 241 combinations of input parameters of multiplicative and additive polynomials, similar to the procedures presented in \citet{Gannon2021, Gannon2022, Gannon2023}. The measured RV has been barycentric corrected. For R21, we recovered a systemic velocity of $5536\pm10$\,km\,$\mathrm{s^{\mathrm{-1}}}$. Sources within the 1\,$R_{\rm e}$ were masked out throughout all of the analysis due to their proximity to R21, while objects outside of the aperture were too faint and too far away from R21 to have any impact. 

\begin{figure*}
\centering
\includegraphics[width=\textwidth]{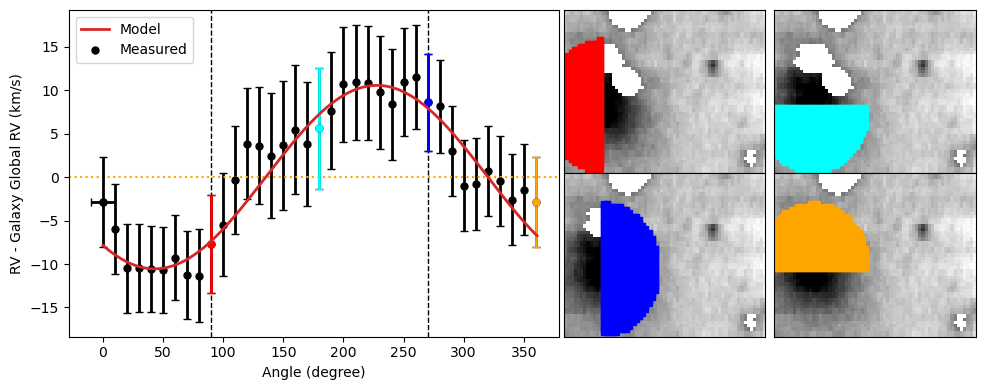}
\caption{Rotation diagram showing the difference between the measured recessional velocity of a given half-region of the galaxy (see coloured regions) and its globally measured RV, plotted as a function of the half’s position angle, with the best fitted model shown in a red line. Dotted lines indicate the position when the halve aligns with the photometric major axis. We show the characteristic uncertainty of the halve position angle only in the first point to avoid overcrowding the plot. For clarity, we additionally display the halves at position angles of 90$^{\circ}$ (red), 180$^{\circ}$ (light blue), 270$^{\circ}$ (dark blue), and 360$^{\circ}$ (orange) in four panels with KCWI cutouts of R21 on the right, with their corresponding locations marked in the rotation diagram. All neighbouring sources in the KCWI image have been masked and are shaded here in white. We find $v_{rot}=10.1\pm6.7$\,km\,s$^{\mathrm{-1}}$, having no correction for disk inclination applied, with a potential hint of rotation axis offset by $\sim30^{\circ}$ to the west of the galaxy photometric major axis at $\sim90^{\circ}$.}
\label{fig:rotation}
\end{figure*}


Next, we performed a simple resolved stellar kinematics analysis of the galaxy to trace its line-of-sight rotation. Due to S/N considerations, we spatially binned spaxels within semicircles of the galaxy, extending out to 1\,$R_{\mathrm{e}}$. To trace the galaxy’s rotation, we offset the halves by 10$^{\circ}$ in a counter-clockwise direction for each iteration. This is similar to the methodology used in inferring galaxy cluster rotation by dividing the projected distribution of cluster member galaxies into two halves, each rotated by an angle on the plane of the sky \citep[e.g.][]{Manolopoulou2017}. In Figure\,\ref{fig:rotation}, each of the black points indicates the difference between the measured recessional velocity of a given half-region of the galaxy and the globally measured RV. We also show four examples of the spatially binned halves of the galaxy at different PAs with their corresponding rotation measure.

We derive a line-of-sight rotation velocity, $v_{\rm rot}=10.1\pm6.7$\,km\,s$^{\mathrm{-1}}$, without correction for any potential disk inclination. The rotation is also further diluted by averaging the velocity measure over semi-circular aperture, thereby representing the lower limit of the inferred galaxy rotation measured within a 1\,$R_{\mathrm{e}}$ aperture. To mitigate the PA averaging dilution, we apply an additional correction factor of $\pi/2$ for the case of semicircular aperture, thus elevating the projected rotation measure to 15.6\,km\,s$^{\mathrm{-1}}$. There may also be a hint of offset of the rotational axis to the west from the photometric major axis by $\sim30^{\circ}$, although this should be treated with caution, as the PA remains poorly constrained due to the galaxy's roundish morphology. For consistency, we applied the same procedure using the BH3 dataset, but extending to a $\sim0.7R_{\mathrm{e}}$ aperture due to S/N considerations. This still yielded a consistent $v_{\rm rot}=9.7\pm4.5$\,km\,s$^{\mathrm{-1}}$, with the same hint of offset in the rotation axis.

\begin{figure*}
\centering
\includegraphics[width=\textwidth]{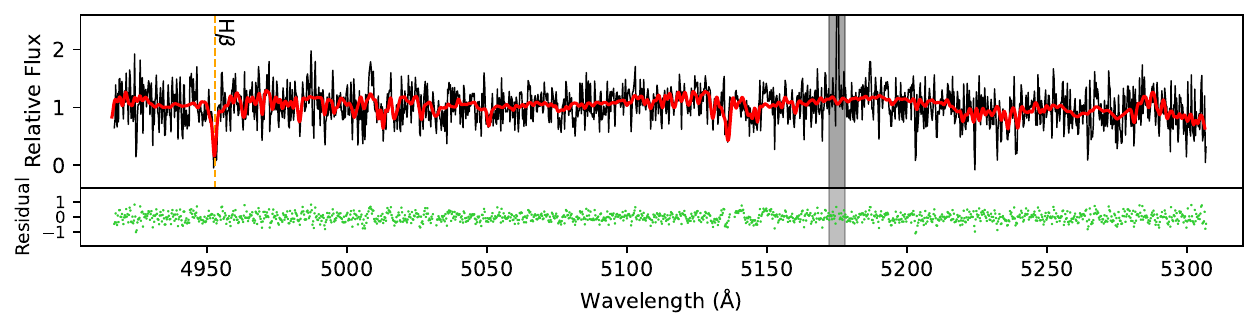}
\caption{KCWI BH3 spectra of R21 extracted over 1\,$R_{\mathrm{e}}$ aperture with a S/N of $\sim$12. We show the observed spectrum (black) with a {\sc pPXF}-based model fit (red), with the vertical shaded band indicating the masked skyline subtraction residual. We also mark the position of strongest absorption feature here, H$\beta$, using the vertical orange dotted line. Our velocity dispersion measurement (see text) is derived from this spectrum.}
\label{fig:bh3_spec}
\end{figure*}

\begin{figure}
\centering
\includegraphics[width=7.8cm]{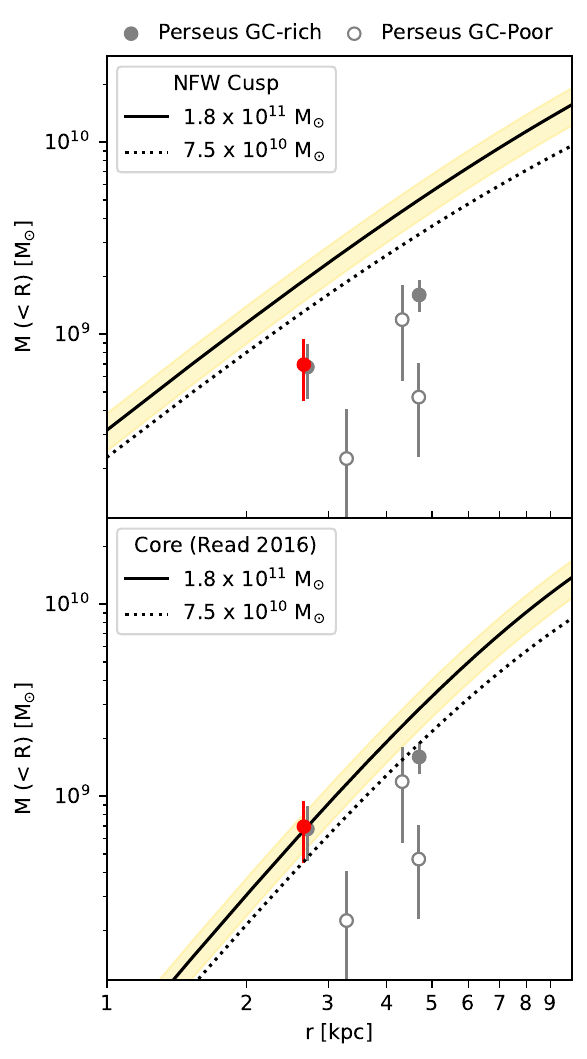}
\caption{Enclosed mass profiles as a function of projected galactocentric radius are shown using a standard NFW cuspy dark matter profile (top panel) and a cored dark matter profile from \citet[ bottom panel]{Read2017}. Mass profiles in solid lines assume a total halo mass of 1.8\,$\times$\,10$^{11}\,\rm M_{\odot}$, derived using the $N_{\mathrm{GC}} - M_{\mathrm{halo}}$ relation from \citet{Burkert2020}, whereas the dotted ones assume a halo mass of 7.5\,$\times$\,10$^{10}\,\rm M_{\odot}$ from the standard stellar mass-halo mass relation from \citet{Behroozi2013}. Yellow shading represents the corresponding error associated with the uncertainty in the GC counts. We included the measured dynamical mass within\,$r_{\mathrm{1/2}}$ for R21 and its associated uncertainty, shown as a red symbol. We also show the measured dynamical masses of other GC-rich (filled grey) and GC-poor (open grey) Perseus UDGs from \citet{Gannon2022} for comparison. Therefore, under the assumption of the GC-based derived halo mass, R21 is likely to reside in a more cored dark matter profile.}
\label{fig:dmmass}
\end{figure}


\subsection{Cluster Infall}


Using the recovered recessional velocity of $5536\pm10$\,km\,$\mathrm{s^{\mathrm{-1}}}$ and a projected cluster-centric distance of 0.34\,Mpc, we place R21 in a position--velocity phase space diagram, which identifies different epochs of infall times of cluster-member galaxies \citet{Rhee2017}. 
For Perseus cluster, we adopted: $R_{\mathrm{200}}=2.2$\,Mpc, $V_{\mathrm{r}}=5258$\,km\,s$^{\mathrm{-1}}$ and $\sigma_{\mathrm{cluster}}=1040$\,km\,s$^{\mathrm{-1}}$ \citep{Aguerri2020}. 
With these inputs, R21 is placed within the sample of galaxies that likely have fallen into the cluster at the earliest times, and is thus classified as an ``ancient in-falling'' galaxy.

We caution against over interpreting this classification, given the statistical nature of number densities in each infall group and the presence of interlopers. Nonetheless, given that the timescales for environmentally driven quenching are $\sim$\,1 to 2\,Gyr \citep[e.g.][]{Fillingham2015}, we could approximate the infall times of R21, as $\rm t_{\rm{inf}}\sim\rm t_{\rm{90}}-1.5$\,Gyr ago \citep[See e.g][]{Anna2018, Anna2023}. This translates to an infall time of $\sim$10.3\,Gyr, compatible with the early infalling classification from the phase-space diagram (see Table\,\ref{tbl:outputs} for values of $\rm t_{\rm{inf}}$ and $\rm t_{\rm{90}}$ from the full spectrum fitting analysis). 




\subsection{Velocity Dispersion}\label{subsec:vdisp}

Next, we measured the global stellar velocity dispersion of R21 using the integrated spectrum extracted from the BH3 data, owing to its higher spectral resolution. The extracted spectrum is shown in Figure\,\ref{fig:bh3_spec}, with the detected absorption line, H$\beta$ indicated. The fitting procedure followed the methodology of \citet{Gannon2021}, and it is identical to the approach used for deriving systemic velocities with {\sc pPXF}, yielding a stellar velocity dispersion within 1\,$R_{\mathrm{e}}$ of $\sigma_{\mathrm{e}}=19.4\pm3.5$\,km\,s$^{\mathrm{-1}}$.

From this velocity dispersion we can further infer the dynamical mass of the galaxy, using the mass estimator equation from \citet{Wolf2010}: 
\begin{equation}
M_\mathrm{dyn} = \; 930 \; \Bigg(\frac{\sigma_{\mathrm{e}}^{2}}{\mathrm{km}^2 \, \mathrm{s}^{-2}}\Bigg) \Bigg(\frac{R_{\mathrm{e,circ}}}{\mathrm{pc}}\Bigg) \; \mathrm{M}_\odot\, ,
\end{equation}

\noindent
with the circularized half-light radius $R_{\mathrm{e,circ}}=R_{\mathrm{e}}\times\sqrt{b/a}$, and the measured line of sight velocity dispersion $\sigma_{\mathrm{e}}$. Despite the presence of galaxy rotation in R21, the measured $\sigma_{\mathrm{e}}$ remains to represent a good approximation of the combined rotational and dispersion moments \citep{Courteau2014}.


Using the measured velocity dispersion of $19.4\pm3.5$\,km\,s$^{\mathrm{-1}}$, $R_{\mathrm{e}}=2.16$\,kpc and $\sqrt{b/a}\simeq 0.92$, we find M$_{\mathrm{dyn}}=9.3\pm3.3\times10^{8}$ M$_{\odot}$ within the 3D half light radius, $\rm r_{1/2}$, which is 2.66 kpc. 

\subsection{Stellar and Halo Mass}\label{subsec:halomass}
For the stellar mass of galaxy, we made use of HST-based photometric fluxes. The corresponding M$_{*}$/L$\rm _{F814}$ ratio from the compiled templates from stellar population fitting (See Section\,\ref{sec:stellarpop}) is 2.05 for F814W, with the $M_{\rm F814W}= -16.5$. This produces a galaxy stellar mass of $3.25\pm0.36\times10^{8}$ M$_{\odot}$.

With the dynamical mass derived, we can further compare our results to the known estimates of total halo mass of galaxy incorporating the well-established GC to halo mass relation \citep{Burkert2020} and the one derived from the stellar mass-halo mass (SMHR) \citep{Behroozi2013}. Using the estimate of 36$_{-8}^{+8}$ GC candidates, we can calculate a total halo mass of 1.8\,$\times$\,10$^{11}$M$_{\odot}$ from the GC-halo mass relation. Subsequently, using the galaxy stellar mass, the total halo mass using the SMHR is 7.5\,$\times\,10^{10}$M$_{\odot}$. We also note of a third approximation using the weak lensing analysis from \citet{Sifon2018}, with an estimated dark matter halo mass of $10^{11.8}$M$_{\odot}$ as the 95\% confidence interval for the upper limit on all UDG halo masses. However, this estimation is well beyond the expected dark matter halo mass here, approaching the masses of $L_{*}$ galaxies. As our dynamical measurement is limited to a measure at a fixed radius, direct comparison with total halo mass models requires an assumption of a dark matter profile. Here, we generated two dark matter profiles constituting a cuspy NFW dark matter halo \citep{Navarro1996}, and a cored dark matter halo \citep{Read2017}. 


To generate the cuspy profiles, we follow the procedure as outlined in the Appendix of \citet{Dicintio2014}, using the halo mass parameters ($\alpha$=1, $\beta$=3, $\gamma$=1) for the NFW prediction. We define our halo concentrations from the halo-mass relationship derived from N-body simulations in \citet{Dutton2014}. The cored dark matter profiles were created by applying a simple transformation to the NFW model described earlier. The methodology follows \citet{Read2017} generating $\rm core$NFW profiles. To set the shallowness of the core in the profile, as in \citet{Forbes2024}, we chose to set the core radius to be 2.75 times the observed half light radius for a maximum core size. The resultant cuspy dark matter profiles are shown on the top panel of Figure\,\ref{fig:dmmass}, while the cored dark matter profiles are presented in the bottom panel.

Mixed core vs cuspy results have been observed in classical dwarf galaxies. Studies of HI kinematics in local dwarf galaxies have reported shallower slopes than those of an NFW cusp, favouring more cored dark matter profiles. On the other hand, we note of the presence of a nucleus in R21, which provides additional some support for a cuspy profile. 
Since merging of GCs is the likely mechanism for forming dwarf galaxy nuclei at this stellar mass \citep{Fahrion2022}, a cuspy profile would be preferred. A cuspier profile leads to a shorter dynamical timescale and faster in-spiraling of GCs \citep{Hernandez1998}, thus increasing the chance of nucleus formation. Recent studies such as \citet{Lipka2024}, examining dwarf ellipticals (dEs) from the VIRUS-dE survey, also found denser, cuspy dark matter cores in non-rotating dEs near the centre of the Virgo cluster. The cored profiles however were found to be in rotating dEs with more recent star formation histories and located in the cluster outskirts.  

Here, comparing the cuspy and cored dark matter profiles with the measured dynamical mass of R21 in Figure\,\ref{fig:dmmass}, our results also indicate that R21 likely resides in a more cored dark matter halo profile under all the aforementioned assumptions. The measured dynamical mass also falls within the measured uncertainty for both the cored dark matter profiles derived from the GC-halo mass relation and the SMHR relation. R21 also complements the sample of five other Perseus UDGs with varying GC richness studied in \citet{Gannon2022}, further reinforcing the observed trend of increased dynamical masses in GC-rich UDGs. Following the GC--halo mass relation, we also note of a slight preference for cored dark matter profiles. Such preference for cored profiles is also noted in \citet{Forbes2025}, where GC-rich UDGs are shown to exhibit better agreement with cored dark matter profiles, based on total halo masses inferred from GC counts (see Figure\,7 of \citet{Forbes2024} for a comprehensive sample of UDGs with N$_{\rm GC}$ > 20). When compared to existing simulations, core profiles naturally arise in feedback-based models such as the NIHAO simulations \citep{Dicintio2017}. The Illustris-dark simulation also shows a preference for core profiles, demonstrating greater susceptibility to tidal heating and stripping for UDG formation \citep{Carleton2019}. This leaves all of these models as plausible formation channels for R21.

\section{Stellar Populations Analysis}\label{sec:stellarpop}

\begin{table*}
	\centering
    \caption{Physical properties of R21 integrated stellar population.}
	\label{tbl:outputs}
	\begin{tabular}{cccccccccc} 
		\hline
		N$_{\mathrm{GC}}$ & M$_{*}$ & M$_{\mathrm{halo}}$ & Age,\;t$_{\mathrm{m}}$ & [M/H] & t$_{\mathrm{90}}$ & t$_{\mathrm{50}}$ & [Mg/Fe] &  $\nabla$log(Age) & $\nabla$log([M/H])\\
         & (10$^{8}\,\mathrm{M_{\odot}}$) & (10$^{10}\,\mathrm{M_{\odot}}$) & (Gyr) & (dex) & (Gyr) & (Gyr) & (dex) &  (Gyr/R$\rm {_e}$) & (dex/R$\rm{_e}$)\\
		\hline
        36 $\pm$ 8 & 3.25 $\pm$ 0.36  & 18 $\pm$ 4 & 10.40 $\pm$ 1.20 & -0.64 $\pm$ 0.12 & 8.88 $\pm$ 1.13 & 11.6 $\pm$ 1.32 & 0.38 $\pm$ 0.25 &  $-$0.024 $\pm$ 0.05 & $-$0.002 $\pm$ 0.014 \\
		\hline
	\end{tabular}
    \begin{tablenotes}
        \item{\textbf{Notes.}}  Estimated GC counts from \citet{Janssens2024} in column 1. Stellar mass (see Section\,\ref{subsec:AgeZ}) and (pre-infall) dark matter halo mass are listed in columns 2 and 3. The halo mass was calculated using the N$_{\mathrm{GC}}- M_{\mathrm{halo}}$ relation from \citet{Burkert2020} with the assumption of one globular cluster per 5\,$\times$\,10$^{9}$\,M$_{\odot}$ of dark matter. The global mass-weighted age ($\mathrm{t_{m}}$) and metallicity are listed in columns 4 and 5. $\mathrm{t_{50}}$ and $\mathrm{t_{90}}$ are the ages when 50 and 90 percent of stellar mass was formed (columns 6 and 7). [Mg/Fe] measure (column 8) based on a line index approach (see Section\,\ref{subsec:alpha}). The age and metallicity gradients computed within 1\,R$\rm_{e}$ are shown in the columns 9 and 10. 
    \end{tablenotes}
\end{table*}

\begin{figure*}
\centering
\includegraphics[width=\textwidth]{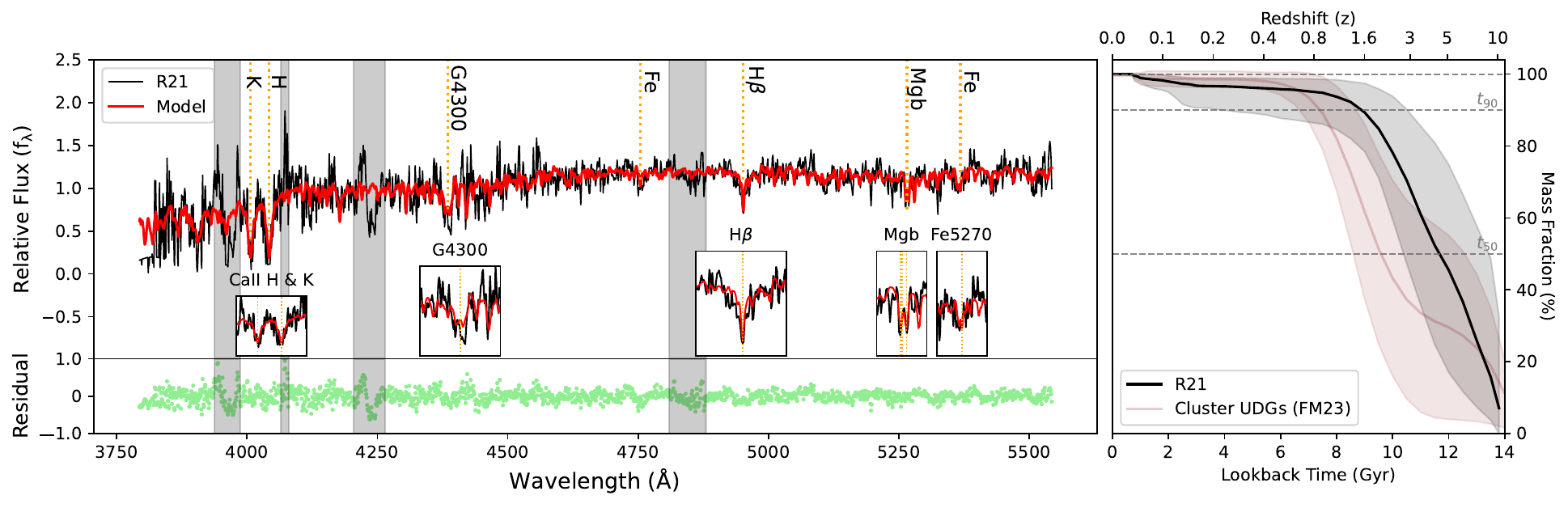}
\caption{Left panel: {\sc pPXF} fit to the spectrum of R21. The black curve show the integrated continuum-corrected spectrum extracted over the radius of 1\,$R_{\mathrm{e}}$ with all neighbouring sources masked out. The red curve indicates the best-fitting model generated by the {\sc pPXF} with the residuals of the fit shown as green points below. Shaded regions indicate masked out portions of the spectra. We also show exerts of zoomed-in parts of the spectrum that contain major detected absorption line features such as the Calcium doublet, G band, H$\beta$, and metallic lines as Mg$_{\mathrm{b}}$ and Fe5270 lines. Right panel: We show the cumulative mass fraction of R21 (in black) as a function of look-back time. Overlaid is the average cumulative mass fraction of UDGs in high-density environments (in red) from the sample of \citet{Anna2023}. Horizontal grey lines mark where 50\%, 90\%, and 100\% of the stars has been formed in the galaxy. The star formation history of R21 thus clearly shows an early and rapid stellar assembly, closely resembling those of UDGs in high-density environments.}
\label{fig:UDGspec}
\end{figure*}

We measured stellar population parameters from the BL spectrum due to the larger number of absorption features contained in its longer wavelength range and the higher S/N of the spectrum ($\sim$26). We performed {\sc pPXF} full-spectral fitting procedure incorporating the  E-MILES templates \citep{Vazdekis2015} with a Kroupa Universal \citep{Kroupa2001} initial mass function and BASTI isochrones. This set of templates provides a fine sampling in metallicities, ranging over $-2.27\leq$[M/H]$\leq0.40$\,dex, and ages ranging from 30\,Myr to 14\,Gyr. All of the stellar population models are scaled solar models. 

We chose to adopt a high-order Legendre multiplicative polynomial for the fit owing to unusual spectral continuum shape (likely caused by imperfections in spectral response calibration). We make use of common normalization technique ($\rm \lambda_{max}-\rm \lambda_{min})/100$\,\AA \,\, \citep[e.g.][]{Gu2018}, adopting a 17th order multiplicative polynomial factor. To ensure that the polynomial minimally affects the overall result of the fitting, we also performed similar fits with polynomials with varying orders of 7 to 20, finding minimal variations in the results, all being within the measured uncertainty. We show the non-regularized fitting median output solutions with 16th and 84th percentiles taken as the 1$\sigma$ uncertainties calculated using 1000 bootstrap fitting iterations \citep{Capperllari2017}. We also performed checks to ensure that the results are independent of the regularization, with bootstrapping procedures providing a similar smoothing of the final SFH output.  
For the alpha abundance measurement, we use a line index approach as outlined in \cite{Anna2023}, see below.

\subsection{Global Stellar Ages and Metallicites}\label{subsec:AgeZ}

To determine the global stellar age and metallicity of R21, we extracted an integrated spectrum by collapsing the spaxels within 1\,$R_{\mathrm{e}}$. All other sources contained within 1\,$R_{\mathrm{e}}$ were masked out. From the spectral fitting, we recovered a mass-weighted stellar age $t_{m}=10.40\pm1.20$\,Gyr with a metallicity [M/H]$=-0.64\pm0.12$\,dex. The continuum corrected R21 spectrum and its {\sc pPXF} fit are shown in the left panel of Figure\,\ref{fig:UDGspec}, with some of the prominent absorption lines marked, such as CaH$+$K, H$\beta$, G4300, Mg$_{\rm b}$ and some Fe lines. Based on the cumulative mass fraction over cosmic time, shown in the right panel of Figure\,\ref{fig:UDGspec}, we calculate the galaxy's age at which 50\% and 90\% of its stellar mass were formed: $t_{50}$ = 10.6 $\pm$ 1.32\,Gyr and $t_{90}$ = 8.88 $\pm$ 1.13\,Gyr.  




\begin{figure*}
\centering
\includegraphics[width=\textwidth]{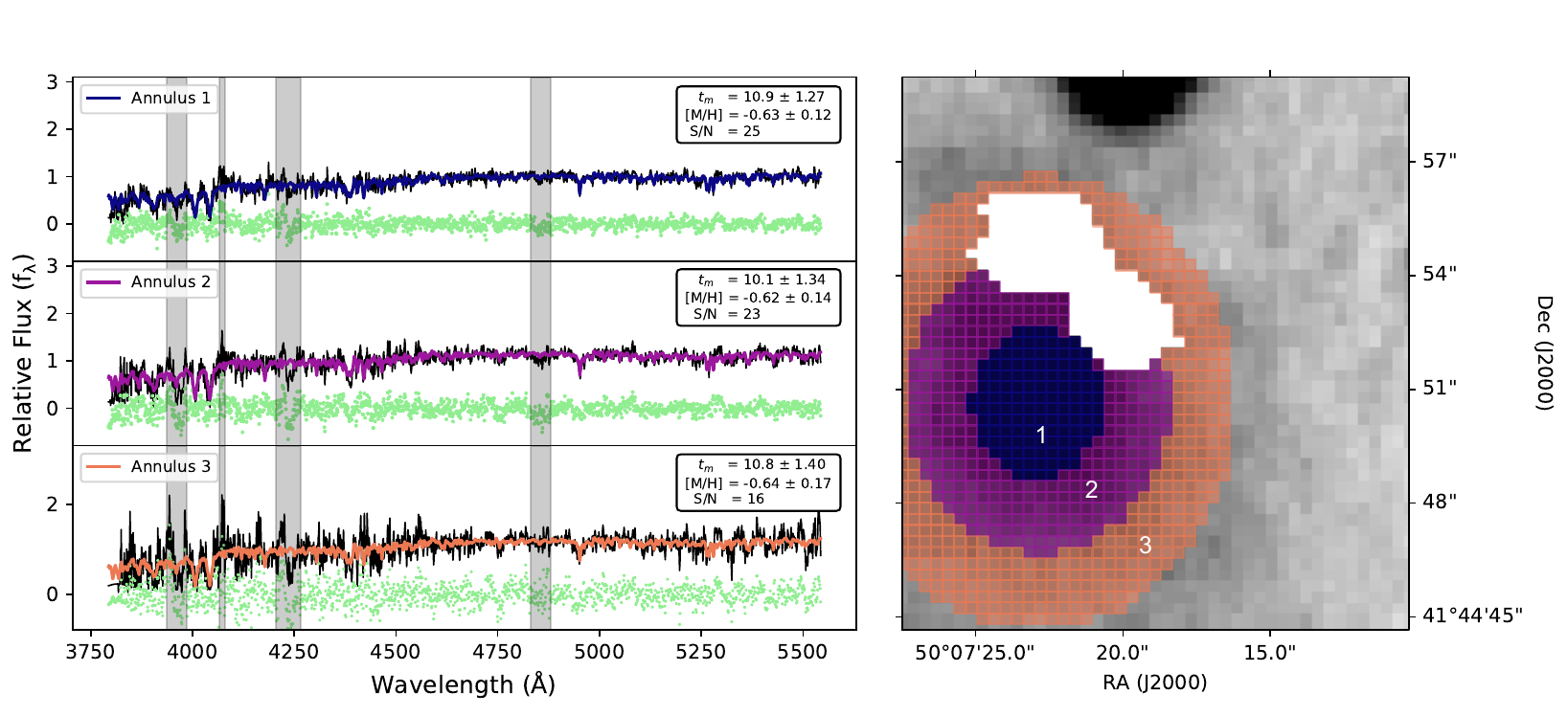}
\caption{Radial stellar population properties of R21. The {\sc pPXF} fits to the spectra for three annular regions of R21 are shown in the left panel, coloured based on the corresponding annular region indicated in the cutout of the KCWI image (right panel; blue, purple, and orange). Neighbouring cluster member galaxies have been masked out as shown by in white colour. The residuals of the fits are shown as green points below each spectrum and grey shaded regions indicating masked parts of the spectrum. We indicate the best-fit mass-weighted ages, metallicities and S/N of spectra on the top right corner of each fit. Overall, we find that the galaxy shows flat gradients, both in age and metallicity.} 
\label{fig:gradientspec}
\end{figure*}

\subsection{$\alpha$-enhancement}\label{subsec:alpha}
To estimate of the $\alpha$-enhancement ([Mg/Fe]), we follow a procedure similar to that used for UDGs \citet{Anna2018, Anna2023}, employing three different methodologies. Below is a brief outline of these procedures:

The first method uses an age-sensitive index (e.g. H$\beta_{o}$;  \citealt{Cervantes2009}) and a metallicity one (e.g. Mg$\rm _{b}$ and $\langle$Fe$\rangle$) to obtain their abundances in the index-index model grid. With these, we measure [Mg/H]$=\rm{Z_{Mg}}$ and [Fe/H]$=\rm{Z_{Fe}}$, which are then used as a zeroth-order approximation of [Mg/Fe], based on the proxy $\mathrm{\rm Z_{Mg} - Z_{Fe}}$ \citep{Vazdekis2015}. This results in 
[Mg/Fe]$\sim 0.33\pm 0.24$\,dex.

The second approach relies on measuring the equivalent width of magnesium (e.g., Mg$\rm _b$) and iron (for e.g., $\langle{\rm Fe}\rangle=0.72\times{\rm Fe}5270 +0.28\times{\rm Fe}5335$) spectral indices. Each is then compared against a grid of stellar population models with different $\alpha$-enhancement ([$\alpha$/Fe]$=0.0$ and [$\alpha$/Fe]$=0.4$).
With this method we can directly measure the [$\alpha$/Fe] ratio by interpolating between the two model grids, resulting in [$\alpha$/Fe] of 0.27\,dex. This value is then converted into [Mg/Fe] using an empirical relation from \citet{Vazdekis2015}, [Mg/Fe]$=0.59\times [\alpha$/Fe], which would yield [Mg/Fe]$\sim 0.46$\,dex.

Lastly, 
we repeat the previous approach but now using the light-weighted age from {\tt pPXF} instead of an age-sensitive index. This provides more orthogonal model grids that make interpolation easier at lower metallicities (e.g. \citealt{Anna2014}) 
With this approach, we obtain [Mg/Fe]$\sim0.35\pm0.25$\,dex.

Here, we report the average value derived from the three methods: [Mg/Fe]$=0.38\pm0.25$\,dex.
\begin{figure}
\centering
\includegraphics[width=\columnwidth]{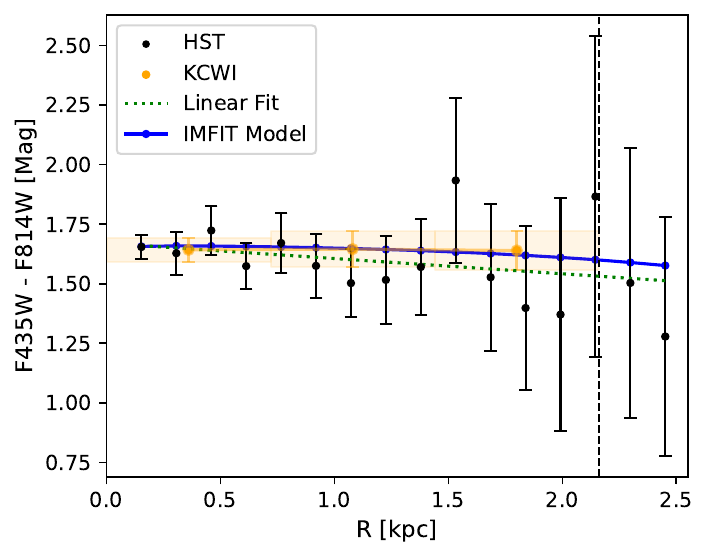}
\caption{Radial F475W $-$ F814W colour profile of R21 from HST imaging. We used 16 annular regions of equal distance apart spanning just further than 1\,$R_{\mathrm{e}}$ (=  2.16 kpc) of the galaxy, which is indicated by the black dashed vertical line. The colours are extracted within the sector with $PA=135$ to $PA=225$ degrees. The resulting colour profile of R21 is shown in black points, with the best linear fit to the data in the green dotted curve having nearly a flat slope of $-$0.023 $\pm$ 0.025\,mag kpc$^{\mathrm{-1}}$. The blue curve indicates the measured colour profile from the IMFIT based model. We overlay the gradient results (in orange) derived from {\sc pPXF} fits to spectra within three annular regions. R21 is consistent with a flat colour profile to 1\,$R_{\mathrm{e}}$.}
\label{fig:gradientphot}
\end{figure}

\subsection{Age and Metallicity Gradients}\label{sec:stellargrad}
Increasing efforts have been carried out to obtain spatially resolved data of UDGs, studying the stellar population properties as a function of galactic radius. Recent findings in \citet{Villaume2022} for DF44, \citet{Fensch2019} for DF2, \citet{Buzzo2025a} for FCC\,224, and \citet{Anna2024} for additional 4 UDGs/LSB dwarfs, have found flat age gradients and flat to positive metallicity gradients, contrary to the bulk of the classical dwarf galaxy population \citep[e.g.][]{Chilingarian2009, Koleva2011, Taibi2022}. Here, we similarly investigate the presence of any age and metallicity gradient in R21. Three radial bins were generated, with the central ellipse having a semimajor radius of 2\arcsec ($\sim 6$\,pixels), with subsequent apertures of two annuli each stepping out by 2\arcsec, extending out to 1\,$R_{\mathrm{e}}$. Each of the spectral bins achieved a S/N$\geq$15\,\AA$^{\mathrm{-1}}$. 

The maximum variation from the global stellar value is 0.4$\pm$1.5\,Gyr in age and 0.02$\pm$0.20\,dex in metallicity, indicating virtually no change in either property. The resultant gradient estimations are $\nabla$log(Age)$\sim-0.024\pm0.05$\,(Gyr/$R\rm_{e}$ and $\nabla$[M/H]$\sim-0.002\pm0.0144$\,(dex/$R\rm_{e}$). The results of the {\sc pPXF} fits along with the spectrum of each annular regions are illustrated in the left panels of Figure\,\ref{fig:gradientspec}. Each of the coloured lines in the right panel of Figure\,\ref{fig:gradientspec} corresponds to the stellar population models associated with the respective annular regions shown in the right panel of the figure. 

\subsection{Colour Gradients}\label{subsec:colourgrad}
We additionally utilized HST imaging to verify if any colour gradient (a strong indicator of a stellar population gradient in the absence of dust gradients) is present in the UDG. We similarly used radial annular bins, each with a radial interval of 0.4\arcsec ($\sim$8\,pixels) extending to 1.15\,$R_{\mathrm{e}}$. The measured colour is computed from a wedge region spanning position angles of $135^{\circ}$\,--\,$225^{\circ}$. This ensures minimal contamination from neighbouring cluster member galaxies in the northern parts of R21. Any remaining sources within the sector were also further masked. The local sky background was calculated using a sky pedestal model with IMFIT\footnote{https://www.mpe.mpg.de/$\sim$erwin/code/imfit/} \citep{Erwin2015} during the fitting of a single S\'ersic model to the UDG in both the F475W and F814W filters. The model sky is subtracted from the images before the flux extraction. We show the final computed colour in each of the annular regions in Figure\,\ref{fig:gradientphot}. The data are fitted with a linear regression function (green line) yielding a negligible slope of $-0.023\pm0.025$\,mag\,kpc$^{\mathrm{-1}}$. 

We also generated a similar radial colour profile from the single S\'ersic models computed for each filter. The model radial colour profile (in blue) is overlaid on top of the observed radial colour profile as shown in Figure\,\ref{fig:gradientphot}. The model fairly resembles the observed profile, with the exception of a few outermost annuli, although this is understandable given the large uncertainties dominated by the sky background. The resulting model has a nearly flat gradient of $-0.01$\,mag\,kpc$^{-1}$, putting it at $\sim-0.06$\,mag over 1\,$R_{\mathrm{e}}$. We also show the colour profile from the spectral fitting results (in orange) from Section\,\ref{sec:stellargrad}, translating the derived spectroscopic-based ages and metallicities from each annular region into photometric colours. We find that the observed photometric colour profile and the one derived from our spectroscopy show good agreement, further reinforcing the trend of virtually flat age and metallicity gradients for this UDG.  


\section{Discussion}\label{sec:discussions}
In the previous sections, we showed through kinematic analysis that R21 resides in the Perseus cluster and exhibits stellar bulk rotation. Its integrated spectrum is also consistent with an old stellar population, characterized by intermediate-low stellar metallicity and elevated $\alpha$-abundances, suggesting an early and rapid assembly of the galaxy, further confirmed by its derived SFH. Building on these findings, we now address the formation origin of R21, specifically examining questions related to the observed flat stellar population gradients, as well as its formation pathways, in comparison with a broader literature sample of UDGs.

\begin{figure}
\centering
\includegraphics[width=\columnwidth]{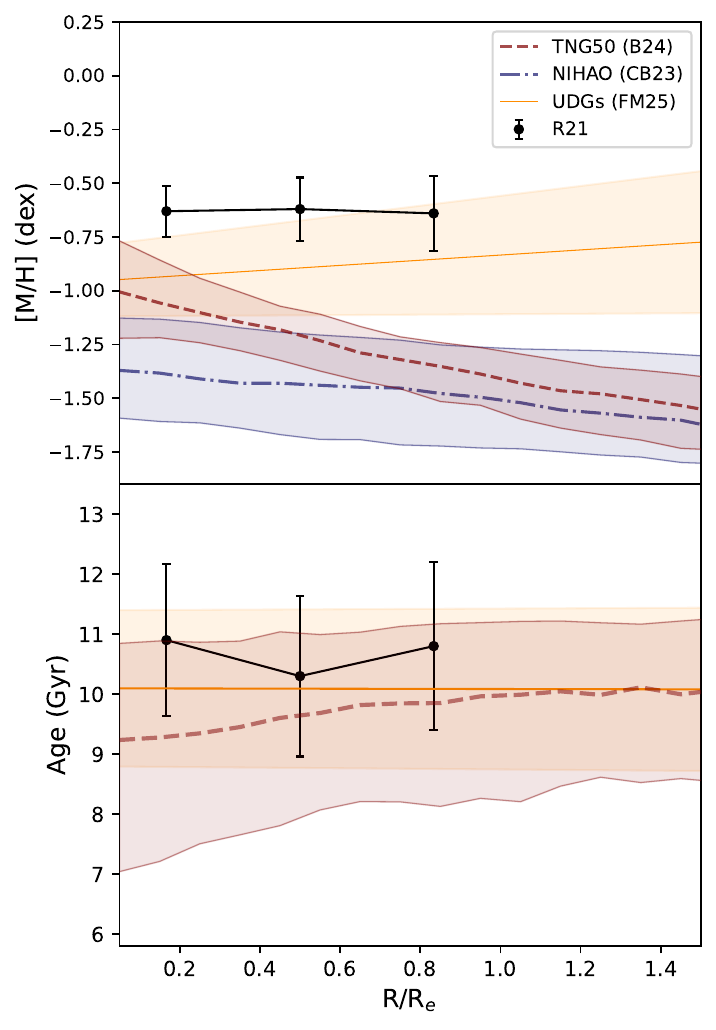}
\caption{Top panel: Metallicity profile of R21 (black solid points) compared to the average metallicity slope of UDGs from \citet{Anna2024} (orange), as well as the average trends from simulated quenched UDGs in cluster from TNG50 \citep[in red]{Benavides2024} and the field ones from NIHAO \citep[in blue]{Cardona2023}. 
We find the metallicity gradient of R21 along with the other observed UDGs clearly diverges from the general predictions of NIHAO and TNG50 simulations. Bottom panel: A similar comparison of age profile for R21 (in black), average age slope of UDGs (in orange) and average trends for simulated cluster UDGs from TNG50 (red). The observed age gradient is overall flat as observed in other UDGs and predictions from TNG50 simulation. } 
\label{fig:gradcomp}
\end{figure}

\subsection{Flat Stellar Population Gradients}
In this work we find similar trends in the derived age and metallicity gradients to previous studies (e.g.DF44 \citep{Villaume2022}, DF2 \citep{Fensch2019}, FCC 224 \citep{Buzzo2025a}, four other UDGs/NUDGes from \citet{Anna2024}), with all exhibiting flat age and flat-to-positive metallicity gradients within 1\,$R_{\mathrm{e}}$. 
Building on Figure\,3 of \citet{Anna2024} (See \citet{Anna2024} for a full discussion regarding comparison between simulated and observed gradients of those UDGs/NUDGes), Figure\,\ref{fig:gradcomp} shows the median age and metallicity gradients from recent simulations, such as TNG50 \citep{Benavides2024} in the top panel, NIHAO \citep{Cardona2023} in the bottom panel (for metallicity only, as the age gradients  are not available), and the average gradient of only ``bona fide" cluster UDGs of \citet{Anna2024} (DF44, DFX1 and DF07) in orange. We overlay our age and metallicity gradient for R21 in black. As in \citet{Anna2024}, we apply a $\sim$$-$0.4\,dex shift to the TNG50 simulation metallicity gradients, matching the metallicities to a given MZR at $\rm log($ $M_{*}/\rm M_{\odot})=9$, while no correction is applied to the field UDGs from NIHAO.  

When compared to the simulations, we find good agreement in the age gradients, which show little to no variation in age as a function of radius. This is expected, given that the TNG50 simulation sample includes only quenched cluster UDGs, which tend to have more truncated star formation histories and a smaller age spread among their stars. Nonetheless, we can clearly see a divergence in metallicity profiles between the quenched cluster UDGs from TNG50 and the observed one from R21. The simulated TNG50 profiles show steeply declining slopes across the entire galaxy radius. The metallicity gradients appear to more sensitively follow the general trend of outside-in formation seen in regular dwarf galaxies, where star formation becomes increasingly centralized over time \citep[e.g.][]{Koleva2011}. This clearly places R21 and other, alike UDGs at the extreme end of the dwarf galaxy population distribution \citep[see e.g.][]{Anna2024} . In contrast to TNG simulation, NIHAO field UDGs exhibit much less steep profiles, despite having much lower metallicities overall. Such flat gradients are found to reside in more rotationally supported galaxies. UDGs that show rotation involved a combination of star formation locus migration and stellar feedback, intermixing newly formed, metal-rich stars in the outer regions with metal-poor stars driven outward by stellar feedback \citet{Cardona2023}. Other flat gradients were also produced in UDGs with more recent stellar assembly, still showing residual star formation at present \citet{Cardona2023}. While the latter explanation is less applicable due to the predominantly older ages of the R21 stellar population, some parallels could still be drawn to the rotating field dwarf galaxies, despite the appreciable differences between field UDGs and cluster ones. Nonetheless, the simulations still struggle to explain all the observed UDG metallicity gradients collectively, where within the observed sample, both rotating and non-rotating UDGs exhibit flat-to-positive gradients.

\subsection{Comparison with Perseus UDGs and LSBs}

\subsubsection{Ages, metallicities and SFHs}
To place R21 in the broader context of UDGs and LSBs, we first take a closer look at its stellar population characteristics. Among the spectroscopically studied Perseus UDGs/LSBs, three (PUDG--R27, PUDG--R84 and PUDG--S74), are considered to be GC-rich 
, hosting 52$\pm$8, 43$\pm$6, and $\sim$30 GC candidates, respectively \citep{Gannon2022,Janssens2024}. They account for $\sim$ 1-2\% of the total stellar mass of each galaxy. 
These UDGs have stellar ages of 8\,--\,11\,Gyr, but two of them have higher metallicities than the bulk of the UDG population \citep{Anna2023}, with [M/H]$=-0.61$\,dex and [M/H]$=-0.4$\,dex for 
PUDG--R27 and PUDG--S74, respectively. These are, in fact, very similar to our findings for R21 both in age and metallicity. All of these UDGs show super-solar $\alpha$ abundances, indicating rapid stellar build-up and early quenching. None of the Perseus galaxies exhibit signs of tidal disturbance indicating they are not the stripped remnants of a more massive galaxy. Overall, these two Perseus UDGs along with R21 lie close or even slightly above to higher metallicities in the present-day dwarf mass--metallicity relation \citep{Kirby2013}. They also fall comfortably within the age--metallicity--alpha abundance parameter space occupied by observed dEs \citep{Anna2023}, further resembling the population of classical dwarf galaxies. PUDG--R84, on the other hand, deviates strongly from that relation, being more compatible with galaxies on the mass--metallicity relation at ${z}\sim2$ \citep{Anna2023}, resembling a failed-galaxy type instead \citep{Forbes2025}. 

In the context of the star formation history, as shown in the cumulative mass fraction plot in Figure\,\ref{fig:UDGspec}, R21 exhibits a rapid assembly history. The calculated lookback time since the buildup of 90\% of the stellar mass ($t_{90}$) is 8.88 Gyr (or equivalently, using the commonly adopted quenching timescale since the Big Bang, indicating a short time of $\Delta t_{90} \sim 4.9$ Gyr), further reaffirming the galaxy’s likely early infall into the cluster environment. This is further supported by the resemblance of R21’s inferred SFH to those of UDGs in high-density environments, as shown in Figure\,\ref{fig:UDGspec}, where it falls well within the intrinsic scatter of that sample. Alongside elevated alpha abundances, these findings are fully consistent with the sample of Perseus and Coma UDGs studied by \citet{Anna2023}, who report a clear trend that UDGs with rapid early star formation histories also host the oldest and most alpha enhanced stellar populations.


\subsubsection{Galaxy and GC colours}
These GC-rich Perseus LSBs that are on the mass-metallicity relation also show small colour difference between their GC population and the main galaxy stellar body. The GCs having a single colour distribution bluer than the remaining stellar body, implying a lower inferred metallicity.  
For the case of R21, the spectroscopic-based age and metallicity of galaxy stars correspond to an HST based colour (F475W$-$F814W) of $\sim$1.65\,mag$\pm$0.07, while the mean GC colour measured in \citet{Janssens2024} is 1.51\,mag$\pm$0.03. This gives the $\triangle$(F475W$-$F814W)$_{\mathrm{UDG-GCs}}=0.14$\,mag$\pm$0.08. Despite being still within the observed uncertainty, there is a hint that the GC population is bluer, consistent with the interpretation of being more metal poor than the galaxy starlight. This can be demonstrated with simple single stellar population (SSP) to colour conversion assuming an old age such as a 10\,Gyr stellar population age. The mean GC colour translates into [M/H]$\simeq-1.2$\,dex, while it is [M/H]$\sim-0.7$\,dex for the stellar body.  An even larger colour difference is observed for PUDG--R27, with $\triangle$(F475W$-$F814W)$_{\mathrm{UDG-GCs}}=0.17$, potentially indicating at least two stellar populations constituting the metal-poor GC population and bulk of the main stellar body of these UDGs. This would not be dissimilar to the classical elliptical dwarfs in Virgo cluster, with a GC to host galaxy colour difference found to be $\sim$0.1\,mag \citep[e.g.][]{Peng2006} using similar HST filters. 

There are also galaxies with more subtle colour differences, such as PUDG--R84, which has a $\triangle$(F475W$-$F814W)$_{\mathrm{UDG-GCs}}=0.08$\,mag, making it the sole GC-rich Perseus UDG with a very metal-poor stellar population ([M/H]$=-1.48\pm0.46$\,dex). This is more analogous to other UDGs like DGSAT\,I, which has nearly identical GC and galaxy star colours \citep{Janssens2022}, and NGC5846$\_$UDG1, with similar ages and metallicities for both GCs and galaxy stars \citep{Muller2020}. This further reinforces the interpretation of R84 being more of the failed galaxy type, 
where the majority of its stellar population originates from tidally disrupted GCs, as outlined in \citet{Forbes2024}. 


\subsubsection{Stellar Rotation}

Another aspect of comparison between these different UDGs is their stellar bulk rotation. The Perseus dwarfs studied to date were all integrated measurements that would not be able to detect rotation, if it were present. Nonetheless, we can still compare our results with those of resolved kinematic studies of LSBs/UDGs in other clusters. For instance,  \citet{Buttitta2025} showed that seven out of their 18 LSBs/UDGs in the Hydra cluster have clear rotation signatures, five of which were along an intermediate, off-axis rotation. They found rotation with amplitudes of $\sim$25--40\,km\,s$^{\mathrm{-1}}$ within 1\,$R_{\mathrm{e}}$, which is stronger than the one inferred for R21. In terms of kinematic support, five of those rotating galaxies were interpreted to be rotationally supported, in contrast to some well studied UDGs like DF2 \citep{Emsellem2019} and DF44 \citep{VanDokkum2019}, which are known to be more dispersion-supported. 

For the case of R21, we can approximate its kinematic support through the crude estimation of the V/$\sigma$ parameter, measuring the ratio between ordered rotation and random stellar orbital motion. As the S/N of our data does not allow for a proper integral-field V/$\sigma$ calculation, we have to rely on global $\sigma$ measured within a 1 $R \rm_{e}$ aperture instead. The global $\sigma$, however, encompasses a combination of rotational and stellar dispersion moments, requiring further accounting for the rotation component (uncorrected for aperture dilution) by $\sqrt{\sigma^{2} - \rm V^{2}}$. Using this and the maximal galaxy rotation at 15.6 \,km\,s$^{\mathrm{-1}}$, we calculate a V/$\sigma$ of $\sim$ 0.95. If we consider the region defining the separation between slow and fast rotators as outlined by \citep{Emsellem2011}, the boundary lies at V/$\sigma$$\sim$0.14 at the ellipticity of R21 at 0.15 ($\epsilon$ = 1 - b/a = 0.15). From this, R21 lies well above the boundary value, regardless of any consideration of the effects of disk inclination, classifying it as a fast rotator, similar to the rotationally supported galaxies in \citet{Buttitta2025}.



\subsection{Proposed Formation Origin of R21}
The clues about the formation R21 are as follows: (i) it contains a substantial population of metal-poor GCs that are likely metal poor based on their colours, consistent with an age of $\sim$10\,Gyr and [M/H]$\approx-1.2$\,dex; (ii) presents an old and early-quenched diffuse stellar population with a low-intermediate metallicity and super-solar $\alpha$ abundance pattern; (iii) shows flat stellar age and metallicity gradients within 1\,$R \rm_{e}$, derived both from spectroscopic data and photometric colours; (iv) shows mild stellar bulk rotation, with a hint that it may be off-axis; and (v) has an inferred massive dark matter halo, with a preference for a cored profile.

With the R21 hosting many GCs, being a prominent characteristic of a failed galaxy, we start by considering this scenario first. At present, no simulations have succeeded in generating failed galaxies, nor have they convincingly predicted UDGs with flat-to-positive metallicity profiles without significant tidal stripping or major stellar bulk rotation. While, the latter could not be ruled out for R21, the presence of flat metallicity gradient and absence of tidal stripping signatures provide partial resemblance to a failed galaxy. Another aspect of failed galaxies relates to the fact that these galaxies hosting many more GCs-per unit starlight, than the general dwarf galaxy population. Indeed they exhibit elevated ratios of GC system mass to galaxy stellar mass with values up to $\sim$10$\%$ \citep{Forbes2025}. Given the old and metal poor stellar population of these galaxies, this suggests that the bulk of the galaxy stellar population consist of tidally disrupted GCs that likely formed in a single star formation event. In the case of R21, it has a more conservative M$_{GC}$/M$_{*}$ of 1.7\%, but is still lying on the much higher side of the distribution of classical dwarfs \citep[e.g.][]{Forbes2018, Chen2023, Buzzo2024}. The high GC counts also translate into the predictions of a massive halo, lying off the SMHR to higher masses. This still remains possible given the dynamical mass measurement of R21, lying within the uncertainty of both predictions from GC-halo mass and SMHR. Nonetheless, a drawback for a failed galaxy interpretation in R21 is the slightly elevated metallicities of its stellar population, similar to those found in two other GC-rich LSBs/UDGs in the Perseus cluster \citep{Anna2023}. The color offset between the galaxy’s GCs and its stellar population casts some doubt on the interpretation where no further star formation occurs after the GCs form. Given the galaxy's metal-rich stars, this argues against the idea of it being primarily composed of disrupted, metal-poor GCs, implying that at least two separate star formation events must have occurred.

Alternatively, we can examine a more conventional formation origin for R21, associating it with a classical dwarf elliptical galaxies (dEs). As mentioned previously, R21 lies on the present-day mass-metallicity relation, resembling those in the dE population studied in the Virgo \citep[e.g.][]{Sybilska2017}, Fornax \citep[e.g.][]{Romero2023}, and Coma clusters \citep[e.g.][]{Smith2008}. The very early infall time would align with the predictions from \citet{Mistani2016, Carleton2019, Carleton2021} and \citet{Sales2020}, suggesting a higher chance of dEs evolving into UDGs primarily due to tidal heating. The elevated [Mg/Fe] abundance also points to an early rapid stellar assembly for R21, which is consistent with the results of \citet{Liu2016} and \citet{Anna2018, Anna2023}, showing that populations of LSBs and early-type galaxies with higher [$\alpha$/Fe] abundances are associated with higher number GCs for a given galaxy luminosity and shorter star formation histories. \citet{Carleton2021} further inferred increased GC formation efficiencies for these galaxies, owing to higher star formation densities at high redshift, potentially explaining the higher GC numbers associated with R21. The difference in colours between the GCs and the R21 stellar body, reflecting a potential offset in metallicities, can be further explained by the separation in their respective formation times. This is consistent with the inference that galaxies experienced two epochs of star formation before $z\sim2$ based on their GC systems \citep[e.g.][]{Brodie2006, Spitler2010}. 


It is not uncommon to find a bimodal metallicity distribution of GCs in early-type galaxies \citep[\& references therein]{Brodie2006}, which suggests the formation of metal-poor GCs from an early starburst, followed by the creation of more metal-enriched GCs along with bulge star population in massive galaxies. In the case of a dwarf galaxy like R21, much of the metal-rich GC formation may have been suppressed, owing to processes like stellar feedback dissipating large molecular clouds, or due to evolution into an environment with reduced gas turbulence \citep[e.g.][]{Elmegreen1997}. This could have favoured a less efficient, extended star formation process that built up the metal-rich stellar population now composing the stellar body of R21, which is not dissimilar to the population of low-mass dEs, dominated by a largely metal-poor subpopulation of GCs and an absence of metal-rich ones \citep[e.g.][]{forbes2005}. We do note that these two star formation events, however, are not reflected in the SFH of R21 (Figure\,\ref{fig:UDGspec}). This is understandable, given the small time step between the two events occurring at more than 10 Gyr ago, which is not resolved by the spectrum modelling. The latter star formation event would go on until ram pressure stripping removes the last reservoirs of gas during R21's infall into the cluster environment, thereby terminating any remaining star formation. This process would also align with the behaviour of other cluster UDGs \citep{Gannon2022, Chilingarian2019}, with additional processes like tidal heating contributing to the larger sizes observed in UDGs \citep{Yozin2015, Carleton2019, Pfeffer2024}. 

The presence of rotation in R21 can also provide some justification for the flatter metallicity gradients, relating back to the examples of rotating field dwarfs from NIHAO. However, a rotating dwarf is not entirely consistent with the prediction of \citet{Pfeffer2024}, who find that rotationally supported galaxies preferentially have lower natal gas pressures, thus forming GCs at lower efficiencies. 
Nonetheless, despite the discrepancy, we still consider an extension of the classical dwarf to be a viable formation scenario for R21.

\section{Conclusions}

In this paper, we have studied one of the most GC-rich UDGs in the Perseus cluster, contributing to the small sample of spectroscopically investigated UDGs in this cluster. Using newly obtained Keck/KCWI observations alongside archival HST datasets, we present the following key findings:

(i) We measured the recessional velocity of R21 ($5536\pm10$\,km\,s$^{\mathrm{-1}}$), confirming its association with the Perseus cluster. We also observed a stellar rotation of $15.8\pm6.7$\,km\,s$^{\mathrm{-1}}$. 

(ii) Using the high-resolution BH3 grating spectra, we measured a global velocity dispersion of $19.4\pm3.5$\,km\,s$^{\mathrm{-1}}$ within 1\,$R_{\mathrm{e}}$. Based on this measurement, we calculated its dynamical mass to be  M$_{\mathrm{dyn}}=9.3\pm3.3\times10^{8}$ M$_{\odot}$ and compared it to core and cuspy dark matter profiles. Assuming the total halo mass derived from GC counts or the stellar mass-halo mass relation, we find that a cored dark matter profile is more consistent with the measured dynamical mass.

(iii) From the higher S/N BL grating spectra, we derived the global stellar population properties of R21. We found that the galaxy consists primarily of an old stellar population (t$_{m}=10.4\pm1.2$\,Gyr) with intermediate metallicity ([M/H]$=-0.64\pm0.12$\,dex) and elevated $\alpha$-abundance pattern ([Mg/Fe]$=0.38\pm0.25$\,dex). These properties suggest the galaxy formed very early on, over a short timescale, compatible with the derived SFH. The intermediate metallicity differs from some known GC-rich UDGs that present overly lower metallicities, but closely resembles that of two other GC-rich UDGs in the Perseus Cluster.

(iv) A resolved study of the stellar population properties in three annular bins revealed a virtually flat age and metallicity gradients. This result is consistent, within uncertainties, with the radial colour profile derived from the HST imaging analysis. The flat age and metallicity profiles align with some of the most recent results from \citet{Anna2024}, who reported flat-to-positive gradients for five UDGs/NUDGes. While there is a good agreement between the observed flat gradients and predicted ones from quenched cluster UDGs from TNG50, we note of a strong contrast between the observed and TNG simulated metallicity gradient results. The observed flat metallicity gradient however does show some parallel to the rotating field dwarfs from NIHAO simulations, which present slightly flatter metallicity gradients.

(v) We predict several possible formation mechanisms for R21, with the preferential scenario linking it to an extension of classical dwarf galaxies, although an alternative failed galaxy scenario could also be possible. Regardless of the formation pathway, we suggest that at least two star formation events are likely at play: the first forming the metal-poor GC population, followed by a more extended second one forming the remaining more metal rich stellar body. The second star formation episode likely terminated, either before or during the galaxy's first infall into the Perseus cluster. 

Nevertheless, to gain a clearer and more comprehensive picture of these cluster UDGs, further spectroscopic studies of both Perseus and other clusters will be essential. Such investigations could provide critical insights into the diverse pathways of UDGs formation and evolution, helping to refine our understanding of these intriguing galaxies.


\section*{Acknowledgments}

We thank the anonymous referee for their constructive feedback and suggestions that helped improve this manuscript. We also thank L. Buzzo, Y. Tang and S. Janssens for their help with observations and R. Davis, A. Di Cintio, S. Cardona-Barrero for insightful discussions. AL acknowledges financial support received through a Swinburne University Postgraduate Research Award throughout the making of this work. DAF, JPB, WJC thank the ARC for financial support via DP220101863.  AFM has received support from RYC2021-031099-I and PID2021-123313NA-I00. AJR was supported by National Science Foundation grant AST-2308390. Some of the data presented herein were obtained at Keck Observatory.
The authors wish to recognize and acknowledge the very significant cultural role and reverence that the summit of Maunakea has always had within the Native Hawaiian community. We are most fortunate to have the opportunity to conduct observations from this mountain. This research also employed observations made with the NASA/ESA Hubble Space Telescope and made use of archival data from the Hubble Legacy Archive, which is a collaboration between the Space Telescope Science Institute (STScI/NASA), the Space Telescope European Coordinating Facility (STECF/ESAC/ESA) and the Canadian Astronomy Data Centre (CADC/NRC/CSA). This research made use of Photutils, an Astropy package for detection and photometry of astronomical sources \citep{Bradley2024}. This research also made use of Montage. It is funded by the National Science Foundation under Grant Number ACI-1440620, and was previously funded by the National Aeronautics and Space Administration's Earth Science Technology Office, Computation Technologies Project, under Cooperative Agreement Number NCC5-626 between NASA and the California Institute of Technology.

\vspace{5mm}


\section*{Data Availability}

The KCWI data presented are available via the Keck Observatory Archive (KOA): https://www2.keck.hawaii.edu/koa/public/koa.php 18 months after observations are taken



\bibliographystyle{mnras}
\bibliography{example} 






\bsp	
\label{lastpage}
\end{document}